# *Time derivatives via interconnected waveguides*

*Ross Glyn MacDonald[1,2], Alex Yakovlev[2] and Victor Pacheco-Peña[1\*]*

[1]*School of Mathematics, Statistics and Physics, Newcastle University, Newcastle Upon Tyne, NE1 7RU, United Kingdom*
[2]*School of Engineering, Newcastle University, Newcastle Upon Tyne, NE1 7RU, United Kingdom*
*\*email: victor.pacheco-pena@newcastle.ac.uk*

**Electromagnetic wave-based analogue computing has become an interesting computing paradigm demonstrating the potential for high-throughput, low power, and parallel operations. In this work, we propose a technique for the calculation of derivatives of temporal signals by exploiting transmission line techniques. We consider multiple interconnected waveguides (with some of them being closed-ended stubs) forming junctions. The transmission coefficient of the proposed structure is then tailored by controlling the length and number of stubs at the junction, such that the differentiation operation is applied directly onto the envelope of an incident signal sinusoidally modulated in the time domain. The physics behind the proposed structure is explained in detail and a full theoretical description of this operation is presented, demonstrating how this technique can be used to calculate higher order or even fractional temporal derivatives. We envision that these results may enable the development of further time domain wave-based analogue processors by exploiting waveguide junctions, opening new opportunities for wave-based single operators and systems.**

# Introduction

In recent years there has emerged a need for new computing paradigms mainly inspired by an increasing difficulty in maintaining the historical rate of computational speedup described by Moore's law[1,2]. In this context, analogue computing exploiting electromagnetic (EM) signals is an example of such promising paradigms. This is due to their potential for high-speed computing (EM waves propagating at the speed of light within the material where the waves travel) and inherent parallelism associated with EM computing techniques[3–5] (where a single structure may be designed to calculate multiple computing processes by exploiting, for instance, different incident polarization, frequency, or angle of the incident signal[6–9]). A remarkable example of analogue computing, and probably one of the founding works in the field, was the *Differential Analyzer* first reported by Hartree in 1935[10]. Such a device was capable of finding the solutions of differential equations through the rotation of differential gears, producing a continuous output solution (i.e., a mechanical computing device). In the context of EM waves, analogue processors are designed to adapt this principle to, instead, compute the solution to equations by applying a mathematical operator directly onto an EM wavefront in either space or time domains[11].

In this realm, different examples of EM wave-based computing structures have been recently reported such as optical networks able to perform computing operations such as matrix inversion[12–15], transverse electromagnetic (TEM) pulse switching with waveguide networks[16–19] and analogue computing with dielectric multilayers[11,20]. Furthermore, the introduction of metamaterials[21,22], artificial media which can exhibit exceptional control over waves in space and time[23–31], has led to the concept of "computational metamaterials" first introduced in 2014 by *Silva et al*[11]. Since then, remarkable examples of metamaterials for computing have been proposed and demonstrated to perform operations such as differentiation and convolution[7,32–37], as well as computing the solutions of more complex operations such as ordinary differential equations and integral equations[6,34,38]. In analogue computing for signal processing, the calculation of derivatives is an especially important task as it enables edge detection, an important first step in any image/signal recognition task[32]. Different EM wave-based analogue processors have been reported performing first order differentiation, in both space and time domains, with examples including structures designed by tailoring the permittivity distribution or reflection/transmission spectra of a metamaterial block/metasurface[9,32–34,38,39]. In practice, this often requires the fine tuning of several design parameters, such as the lengths of dielectric layers in a multilayer structure or the permittivity of a pixel in a 2D grid[9,11,20]. To achieve this goal, various design techniques have been recently applied and demonstrated such as advanced optimization and inverse design [40,41] and also machine learning approaches[20,42,43].

Inspired by the importance of derivatives for computing (and in particular for analogue computing using EM waves) in this work we propose and study a simple device capable of calculating temporal derivatives. The structure is carefully engineered by exploiting interconnected parallel plate waveguides and stubs as transmission lines (TLs). The physics behind the proposed design is presented in detail and the structure is evaluated both theoretically and numerically using the commercial software CST Studio Suite®, demonstrating an excellent agreement between them. As it will be shown, the proposed EM wave-based structure for the calculation of temporal derivatives can be adjusted and optimized by simply modulating one of its parameters (such as the length of the stub waveguides) and can be engineered to work either in transmission or reflection mode, enabling full flexibility in its design. As examples, the proposed structure is implemented to calculate the temporal derivative of different input signals such as sinusoidally modulated Gaussian signals (modulation frequency of 8 GHz) and even arbitrary temporal functions. These findings may lead to the development of other waveguide network-based analogue signal processors capable of performing mathematical and computational operations in the time domain. We envision that such devices may see applications in computing scenarios where operations are applied to large or continuous data sets, such as audio and image recognition with the latter working at frequencies from acoustics, microwaves up to the optical regime.

**Results**

**Theoretical approach**

To begin with, let us first discuss the key aspects involving the operation of an analogue differentiator, as illustrated in Fig 1a. Here, a hypothetical *processor* (grey block) performs the differentiation operation onto the envelope of an incident signal (applied from the left) and returns its solution at its output (right-hand side). As is known, differentiation on a time domain signal $g(t)$ is represented in the frequency domain as a multiplication between the spectrum of the input signal $G(f) = \mathcal{F}\{g(t)\}$ (with $\mathcal{F}$ representing the Fourier transform, and $f$ as the frequency in Hz) by a factor $2\pi i f$ [20,32], representing the transfer function of the differentiator. In this manuscript, all the calculated functions for $G(f)$ from a temporal signal $g(t)$ are normalized to be bounded in the range $0 - 1$; i.e., $G_{norm}(f) =$ $G(f)/\max[G(f)]$ (this is due to the implementation of passive materials in our designs). Now, for signals modulated by a carrier frequency $f_0$, the factor $2\pi i f$ (transfer function) is simply shifted to $2\pi i(f - f_0)$[44], as observed in the set of Fig. 1a. From this, it is clear that if one wants a hypothetical structure to be able to perform temporal differentiation, its transfer function should be able to emulate $2\pi i(f - f_0)$. Importantly, the ideal transfer function of a differentiator $[2\pi i(f - f_0)]$ can produce values larger than one. Again, as we make use of all-passive materials, the magnitude of the transfer function should be bounded between 0 and 1[44]. To account for this, the ideal transfer function $2\pi i(f - f_0)$ is also normalized to be bounded within 0−1 such that the output signal in the frequency domain from the designed structures will have the same range of values as the normalized ideal/theoretical derivative, only differing from the true values by a normalization factor[11,20,38]. With this normalization, a hypothetical device will operate as a first order differentiator if its transfer function resembles a linear and symmetric V-shaped dip centered around $f_0$ (as described above and shown in Fig. 1a).

Now, to design a structure that can emulate the V-shape of the required transfer function in the frequency domain, one can exploit TL techniques (such as filter design) to tailor the transfer function at will. Based on this, in this work we exploit a set of parallel plate stubs connected to a pair (input and output) of waveguides at a central waveguide junction as schematically shown in Fig. 1b. As in our previous works[16,18,19] we consider two types of waveguide junction with parallel plate waveguides connected in either series or parallel configuration. The full details of the splitting and superposition of signals at these junctions can be found in[16,18,19], here we present the basic concepts for completeness. When the characteristic impedance of each waveguide is the same and the cross-section of the junction is small compared to the incident wavelength[45], the splitting of the incident signal after passing the junction will be described by the following scattering matrices:

$$\boldsymbol{A}_{Parallel} = -\boldsymbol{I} + \gamma \boldsymbol{J} \tag{1a}$$

$$\boldsymbol{A}_{Series} = \boldsymbol{I} - \gamma \boldsymbol{J} \tag{1b}$$

where $\gamma = 2/N$ is the transmission coefficient of the junction, $N$ the number of connected waveguides, $\boldsymbol{I}$ and $\boldsymbol{J}$ are the identity and all-ones matrices, respectively. In this manuscript, we will focus on using parallel junctions, however, the same approach could also be exploited using interconnected waveguides in a series configuration (as this discussed in the supplementary materials section S2).

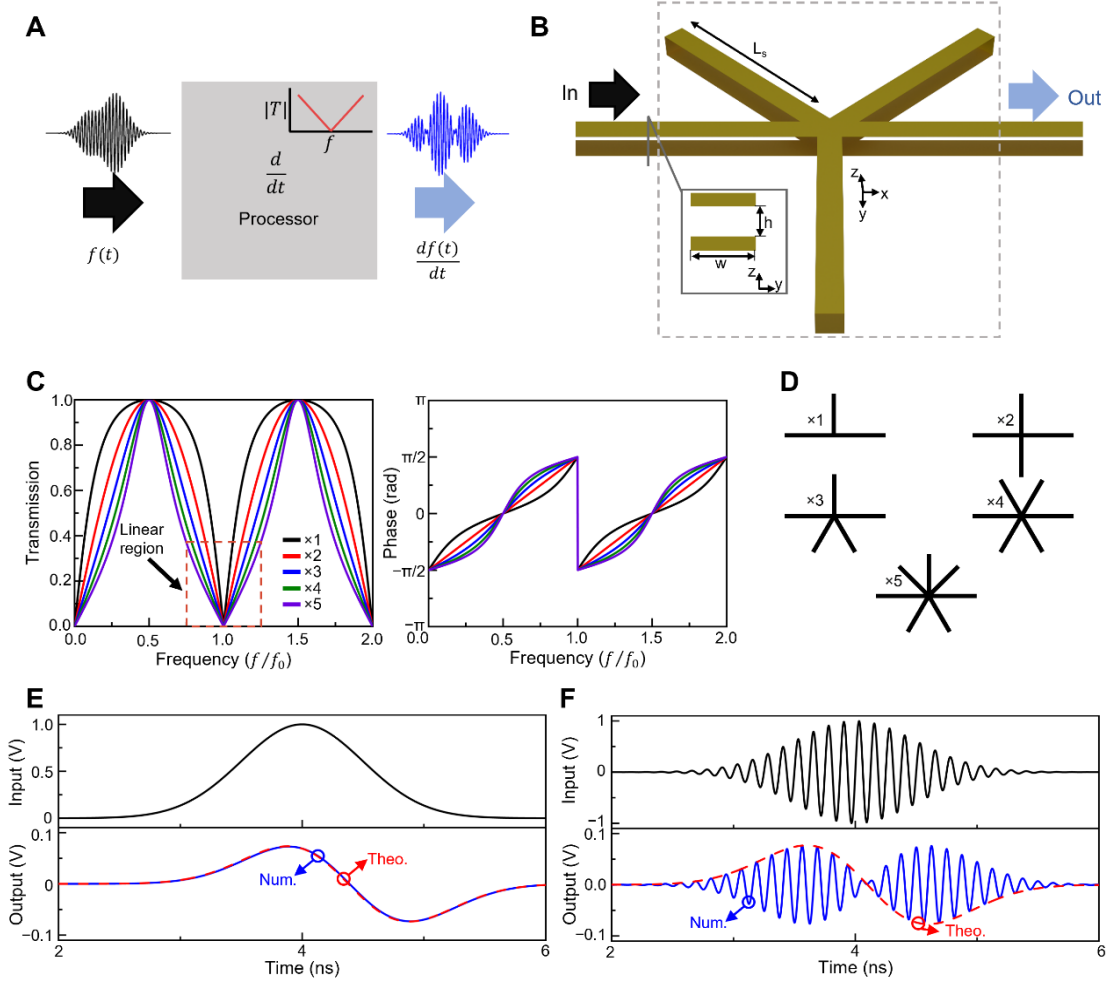

**Fig. 1| Analogue differentiator in the time domain. a**, Block diagram of an analogue differentiator performing first order differentiator onto the wavefront of an incident temporal signal. **b**, Schematic representation of a temporal differentiator designed using three closed-ended stubs connected to input and output waveguides. The waveguides are parallel plates and are connected in a parallel configuration. **c**, Magnitude (left) and phase (right) of the transmission coefficient corresponding to the transfer function of a temporal differentiator using different number of stubs (from 1 to 5) being connected to an input and output waveguide. Here we consider that all the waveguides have dimensions $h = w = 0.0267\lambda_0$. **d**, Schematic representation of the configurations studied in panel **c**, considering different number of closed-ended stubs connected to the waveguide junction. **e, f**, Time domain numerical results of the differentiator presented in **b**, using three waveguides (length $L_s = 0.5237\lambda_0$) connected stubs when excited using an unmodulated (top panel in **e**,) and modulated (8 GHz, top panel in **f**,) Gaussian respectively. The numerical results of the output signals are shown at the bottom panels of **e, f,** (blue lines) along with the theoretical values of the time derivative of the envelope of the incident signal (dashed red line).

As shown in Fig. 1b, we make use of an input (left) and output waveguide (right) interconnecting a network of stubs. With this configuration, the incident signal coming from the left waveguide will split after passing the junction, creating signals traveling towards all the interconnected waveguides. The purpose of the stub waveguides is then to feed these "copies" (scattered signals traveling through them) back into the junction with a small temporal delay (compared to the temporal duration of the incident signal, as it will be discussed below) which can be controlled by the length of the stub waveguides ($\Delta t = 2L\sqrt{\varepsilon_r \mu_r}/c$, with $\varepsilon_r$ and $\mu_r$ as the relative permittivity and permeability of the waveguide filling material, and $c$ is the speed of light in a vacuum), as expected. Here, vacuum is used as the filling materials ($\varepsilon_r = \mu_r = 1$).

We initially consider a pair of closed stubs (terminated with perfect electric conductor, PEC) connected to the input and output waveguide at a parallel junction, such that there is a total of four waveguides connected at the junction. From Eq. 1a, the scattering matrix of this junction is given by $-\mathbf{I} + (1/2)\mathbf{J}$, with $\gamma = 2/4$ i.e. $N = 4$. Based on these values, when an incident signal arrives at the junction from the input (left) waveguide, it will split into four signals of equal magnitude (one traveling along each of the waveguides) which will propagate away from the junction. Specifically, if an incident signal $x_{in}(t)$ has an amplitude of $x_{in}$, from Eq. 1a, a portion of this signal, with an amplitude $x_{in}/2$ will be transmitted to the output waveguide, a copy (also with amplitude $x_{in}/2$) will be transferred to each of the stubs, and a reflected signal with amplitude $-x_{in}/2$ will be excited along the input waveguide; we call this process as the "*first split*". Now, when the signals traveling within the stubs arrive at the metallic terminated ends, they will be reflected with inverted polarity due to the PEC boundary[44]. These reflected signals will travel within the stubs and will arrive back at the junction (after a time delay $\Delta t$ due to the travel time within the stubs, as described above) where they once again will split into four signals as described by Eq. 1a (we call this as the "*second split*"). The superposition of all signals after the second interaction (*second split*) at the waveguide junction will cancel the signals towards the metallic-ended stubs[16,17,19,44] and will leave only two signals propagating away from the junction (one along the input and one along the output waveguides, respectively) both with an amplitude of $-x_{in}/2$, being delayed by a factor $\Delta t$, as explained above. In other words, the signals traveling towards the input and output waveguides after being reflected from the stubs are scattered at the junction and are defined as $-x_{in}(t - \Delta t)/2$. Interestingly, as the incident signal $x_{in}(t)$ is still being applied from the input (left) waveguide, when the *second split* occurs $x_{in}(t)$ will also split and create again four waves after passing the junction (similar to the *first split* as described above). Hence, as the process of splitting repeats, two new signals will travel along the input/output waveguides away from the junction due to the split of $x_{in}(t)$ (with an amplitude of $-x_{in}/2$ and $x_{in}/2$, respectively) which will then interact with the delayed signals created at the *second split* [coming from the metallic-ended stubs $-x_{in}(t - \Delta t)/2$]. Therefore, one can apply again the principle of superposition to show that the signals traveling along the output [$y_{out}(t)$, right] and input [$y_{in}(t)$, left] waveguides, respectively, are the summation of all the signals produced due to the split of the new $x_{in}(t)$ and those delayed signals created by the *second split*, mathematically described as follows:

$$y_{out}(t) = \frac{1}{2}[x_{in}(t) - x_{in}(t - \Delta t)] \tag{2a}$$

$$y_{in}(t) = -\frac{1}{2}[x_{in}(t) + x_{in}(t - \Delta t)] \tag{2b}$$

Interestingly, Eq. 2a indeed resembles the well-known first order finite difference equation[46]:

$$\frac{dx}{dt} = \lim_{a \to 0} \frac{x(t) - x(t-a)}{a} \tag{3}$$

where it is clear how the output equation [$y_{out}(t)$] from Eq. 2a is similar to Eq. 3, only differing by a constant. Due to this, note that for small values of $\Delta t$ (such that the variation in the envelope within a temporal range $\Delta t$ around $t$ may be approximated by a first order Taylor series[46]), the observed output signal will conform to the shape of the first derivative in the time domain. Moreover, as the transfer function (frequency domain) of the differentiator operation from Eq. 3 has a linear V-shape (as explained before), the transfer function of the waveguide network from Eq. 2a will also have a linear V-shape near the frequency of modulation of the incident signal. Importantly, note how the description above has been focused on the "amplitude" of the signals. However, this technique is general and can indeed be applied to incident modulated signals (as it will be shown below). In this case, however, a key factor is the time delay variable $\Delta t$ as it needs to be engineered such that it should ensure that the signals scattered by the junction after the *second split*, are 180° out of phase (i.e., $-x_{in}(t - \Delta t)/2$, as explained above) with the signal transferred to the output waveguide due to the split (new *first split*) of the new incoming, i.e., [$x_{in}(t)/2$] (as described by Eq. 2a). For instance, for parallel plates waveguides filled with air, this condition is fulfilled when the length $L_s$ of the closed (metallic-ended) stub is $L_s = L_{closed} = \lambda_0/2$ (with $\lambda_0$ as the wavelength of the modulation frequency of the incident signal). Note that this could also be done using open-ended waveguides where the condition will be fulfilled when $L_s = L_{open} = \lambda_0/4$[44,47].

**Multiple-interconnected waveguides**

In the previous section, we considered four interconnected waveguides (one input, one output and two stubs). It is also possible to tune the required V-shaped transfer function to meet the needs of specific tasks, such as to control the bandwidth of the differentiation operator being emulated by the network of waveguides. To this end, the transfer function (for instance the transmission/reflection coefficient) of a junction of $N$ waveguides can be parameterized by $M = N - 2$ stubs connected at the junction considering the length of each individual stub, $L_{sj}$, with $j$ representing the stub numbers (from one to $M$) and $\Gamma_{j,\pm 1}$ the reflection coefficient of the individual stubs (again $j$ meaning the stub number, and $\pm 1$ denoting a closed-, $-1$, or an open-, $+1$, ended stub, respectively)[44]. A full mathematical derivation of the transmission and reflection coefficients for an arbitrary combination of parameters can be found in the supplementary materials section S1. In the simplified case where all stubs are identical ($L_{sj} = L_s$, $\Gamma_{j,\pm 1} = \Gamma$) the transfer function can be written as

$$T(f, M, L_s, \Gamma) = \frac{2}{M+2}\left(1 + \frac{2M\Gamma e^{i4\pi f L_s}}{M+2-[M-2]\Gamma e^{i4\pi f L_s}}\right) \quad (4)$$

Using Eq. 4, the transfer function (magnitude and phase of the transmission coefficient in our case) for one to five identical closed stubs of length $\lambda_0/2$ is presented in Fig. 1c along with the schematic representations in Fig. 1d for completeness. As it is shown, the magnitude of the transfer function for all the designs is approximately linear (V-shape) near the normalized frequency $f/f_0$, a required feature if one wants to emulate a differentiation operator as detailed in the previous section. This performance can also be confirmed by looking at the phase discontinuity[48] from the right panel of Fig. 1c which occurs at $f/f_0$. Now, as shown in Fig. 1c, by varying the number of connected stubs at the junction, the spectral width of the linear region around $f_0$ was maximized when three stubs are implemented.

To further evaluate the proposed differentiator using TL techniques, we carried out numerical studies using the time domain solver of the commercial software CST Studio Suite® where full-wave simulations were performed of the structure shown in Fig 1b (i.e., three interconnected closed stubs as the best results of the transfer function from Fig. 1c). Further details of the simulation setup can be found in the methods section below. We consider a Gaussian input signal, both unmodulated and modulated at 8 GHz, as shown in the top panels from Fig. 1e and Fig. 1f, respectively. In both cases, the standard deviation of the Gaussian signal in the time domain was σ = 0.50 ns. The numerical results were compared with the analytically calculated derivative of the envelope of the incident signal (Gaussian un/modulated function) and the results are shown in the bottom panels from Fig. 1e and Fig. 1f. As observed, an excellent agreement between the analytical and numerical results is obtained, demonstrating how, as the designed network of waveguides emulates the transfer function of the differentiator operator in the frequency domain for a derivative in the time domain (V-shape transfer function), it can be used to calculate the temporal derivative of the envelope of incident temporal signals. As detailed above, here we focus our efforts on closed stubs, examples of open stubs, series junctions and combinations of them are shown in the supplementary materials section S2 for completeness.

**Differences between real and ideal scattering**

The transfer functions shown in Fig. 1 were calculated using TL theory by assuming the perfect splitting of the incident signals at the waveguide junction[45,49]. This perfect splitting, as shown in Eq. 1, also considers that the cross-section of all the waveguides is infinitely small or small enough to enable the neglection of fringing fields appearing at the junction. A schematic representation of this perfect splitting behavior is presented in Fig. 2a where an incident signal is equally scattered between all connected waveguides following Eq. 1a. However, as mentioned above, this is an approximation which is only valid for small cross-sections compared to the size of the incident wavelength[45]. Hence, it is important to study the impact of non-ideal scattering on differentiator performance.

Here, two main sources of non-ideal performance can be identified: The first arises from the finite cross-sectional area of the waveguides, which leads to a non-zero junction size. Qualitatively, this enables incident signals to take a shortened path through the junction to the adjacent waveguides, when compared to the ideal splitting model (which considers an ideal zero junction size). This concept is demonstrated in the left panel of Fig. 2b where an incident signal from the left waveguide may travel the reduced red path instead of the ideal green path considered in the theoretical calculations. This performance will translate into a reduction of the effective length of the waveguide stub connected to the junction, resulting in a shift of the transfer function of the device as seen in the leftmost panel of Fig. 2c. In this panel, the magnitude of the transmission coefficient is shown considering a design with two stubs using the theoretical calculations from Eq. 4 (red dashed line) and the numerical simulations using three-dimensional waveguides with $w = h = a = 1$ mm ($0.0267\lambda_0$). Here the parameter $a$ will then be used as a scaling parameter that accounts for a change of the cross section of the waveguides. The frequency shift, represented by the ratio between the numerical and theoretical frequency where the transmission coefficient is almost zero ($f_0$), as a function of the dimension $a$ of the waveguides is shown in the second panel of the same Fig. 2c, confirming that in the limit when $a << \lambda_0$ the frequency shift of the minimum of the numerically calculated spectrum ($f_{min}$) is negligible (with respect to the frequency of the theoretical minimum $f_0$ i.e. $f_{min} \approx f_0$). However, one can compensate the effect of the non-zero junction size by increasing the length of the stubs. To do this, the chosen increase of length must match the total reduction in path length produced by the non-zero junction size, as described above. This is demonstrated in the third panel of Fig. 2c where the shift in the frequency at which the minimum in the transmission coefficient occurred $|\Delta f| = |(f_{min,num} - f_{min,theo})|$ (with $f_{min,num}$ and $f_{min,theo}$ being the frequency at which the transmission coefficient minimum occurred in the numerical simulation and theoretical calculations, respectively) is presented for a range of target frequencies and added lengths $\Delta L$ (normalized with respect to the scaling parameter $a$, $\Delta L/a$). In this study, the target frequency refers to the frequency at which the theoretical minimum of the V-shape of the transfer function appears, assuming perfect splitting and no added length to the stubs; in other words, the target frequency is the modulation frequency of the incident signal $f_0$). The white dashed line shown in this panel represents the amount of normalized added length which minimizes the frequency shift. For instance, the frequency shift in the differentiator with an $f_0 = 8$ GHz target frequency and two stubs with dimensions as those of the example from Fig. 2c, leftmost panel ($a = w = h = 0.0267\lambda_0$) was minimized by a stub length increase of $\Delta L = 0.0227\lambda_0$. The transmission coefficient of this structure after adding the length $\Delta L$ is shown in the rightmost panel of Fig. 2c where it is observed how the numerical simulations with the realistic waveguide are now in good agreement with the theoretical calculations using the TL technique.

The second reason for non-ideal splitting is due to the *effective* spatial asymmetry between the stubs connected at the waveguide junction. For instance, apart from junction size, the reduced path through the junction will also vary with the angle at which the waveguide is connected to the junction, as is schematically shown in the right panel of Fig. 2b. As observed, when multiple stubs are connected to a single junction, asymmetry between the angles of the connected stubs may produce different path lengths through the junction to the individual connected stubs. This will produce a phase mismatch between the signals reflected into the junction from the different stubs. The effect of this onto the transfer function of the device (transmission coefficient in our case) can be studied by looking at the results shown in the leftmost panel of Fig. 2d where the magnitude of the transmission coefficient is shown as the angle between two stubs connected at a 4-waveguide junction (as in Fig. 2b) is varied from $\theta = 0°$ (ideal scenario) to $\theta = 25°$ or $\theta = 45°$, as examples. Here we consider a design with $a = w = h = 1$ mm ($0.0267\lambda_0$) waveguides and stubs with a length $L_s$ of $0.5237\lambda_0$ (for $\lambda_0 = 37.5$ mm, again for a frequency $f_0$ of 8 GHz) measured from the center of the waveguide junction to the metallic-terminated end of the stubs. As it can be seen, as $\theta$ is increased, the linear V-shaped transmission coefficient is distorted due to the phase mismatch between signals from the two stubs. To quantify this distortion, we calculated the root mean squared error (RMSE) between the numerical results of the transmission coefficient for various stub angles (ranging from −60° to 60° with a step of 5°) and an ideal linear V-shaped transfer function centered at $f_0$ ($T_{ideal} = C|f - f_0|$) where $T_{ideal}$ is the transmission coefficient of the ideal function and $C$ is its corresponding scaling constant after it has been normalized to be bound between 0 and 1 within the desired frequency range. The calculated RMSE values are shown in the second panel of Fig. 2d. From these results, the distortion induced by the phase mismatch due to the different angles of the stubs is symmetrical around 0° with distortion increasing when increasing $\theta$, as expected. From a path length perspective, this symmetry can be understood by considering the *first* and *second* split of the signal at the junction: when the *first split* takes place, the incident signal will observe a rotated stub with an angle of $90° + \theta$ (angle between the input and stub waveguides). For the *second split* (after the signal have been reflected by the metallic end of the stub) the signal traveling towards the junction will see an angle of $90° - \theta$ (between the stub and output/input waveguides). Due to this, the distortion in the transmission coefficient will be symmetrical around $\theta = 0°$ as the combined path difference of the *first* and *second split* will be the same for positive and negative angles (only changing the order in which the path differences are observed). This distortion (here measured using the RMSE as discussed above and shown in Fig. 2c) in the transmission coefficient can also be overcome by increasing the length of the rotated stub to compensate for the reduction in its path length. An example is shown in the third panel of Fig. 2d where it is observed how the numerical results of the transmission coefficient for a pair of $a = w = h = 1$ mm ($0.0267\lambda_0$) stubs with one of them being rotated with an 20° angle difference is affected by the added length. With this in mind, a compensation length is added to the end of the rotated stub until the calculated distortion is minimized. For the 25° case, this occurred using an extra length of $\Delta L = 0.6$ mm ($0.0160\lambda_0$) with a target frequency of 8 GHz. For completeness, the required $\Delta L$ to minimize the distortion of the transmission coefficient as a function of the rotation angle of one of the stubs is shown in the fourth panel of Fig. 2d. As observed, no $\Delta L$ was required in the range from −15° to 15°, meaning that experimental/fabrication errors will be negligible as long as the rotation angle of the fabricated waveguide stub does not exceed these values.

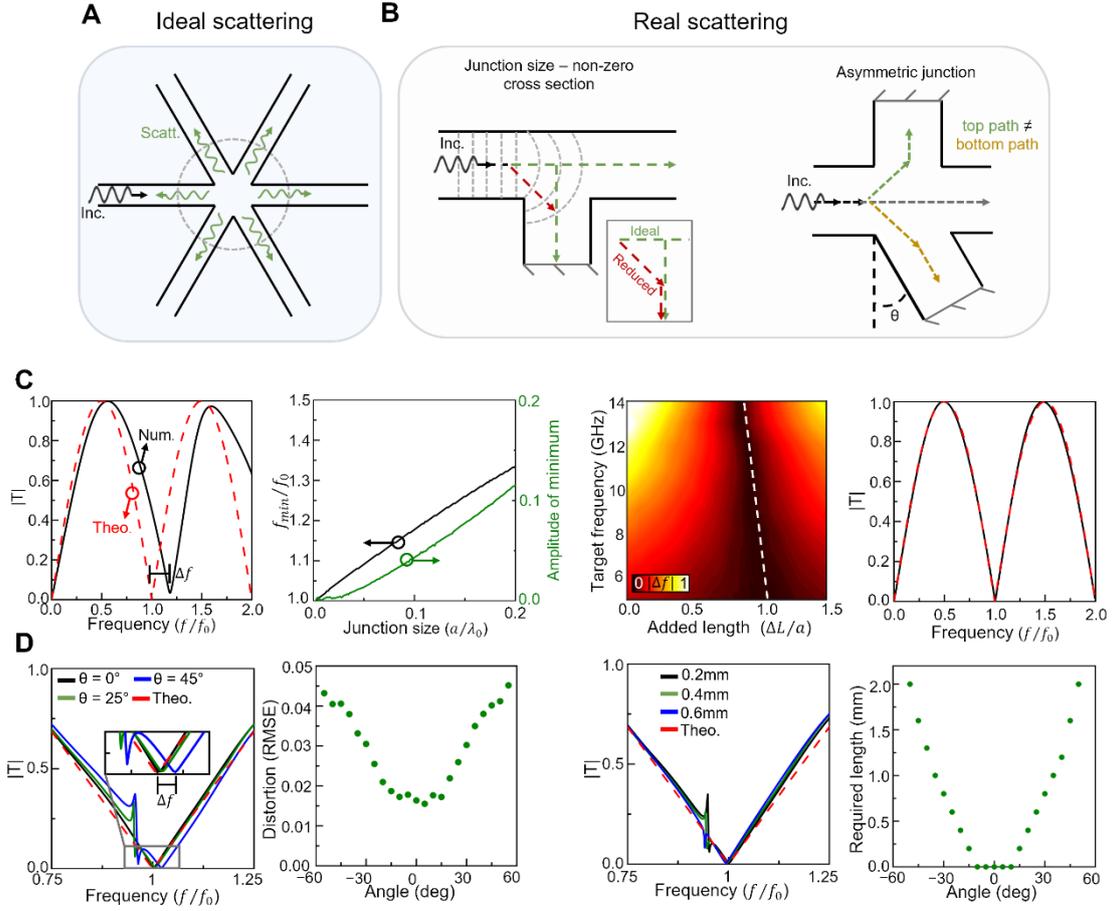

**Fig. 2| Effects of imperfect scattering on the performance of the temporal differentiator. a**, schematic of perfect splitting (in the limit $a \ll \lambda_0$) emulated by the radiation of a dipole. **b**, Non-perfect splitting associated with the non-zero size of the junction cross-section (left) and the potential angular asymmetry of the junctions (right). **c**, and **d**, numerical results showing the impact of the cases shown in **b**, respectively. We consider a temporal differentiator designed with two PEC terminated stubs and waveguides cross-section size $w = h = a = 0.0267\lambda_0$ ($\lambda_0$ as the wavelength in free-space at frequency of 8 GHz). **c**, (left panel) Numerical (black) and theoretical (red) magnitudes of the transmission coefficient for a two-stub differentiator with $L_s = 0.5\lambda_0$. These results show the effect of non-zero junction size on the ideal perfect splitting. (Second panel) Calculated frequency shift between the numerical and theorical minima of the transmission coefficient (black) along the amplitude of the numerical minimum as a function of the junction scaling parameter $a$. (Third panel) Magnitude of the frequency shift (between the numerical and theoretical minima) as a function of the target frequency and the added length $\Delta L$ normalized with respect to the scaling parameter $a$ ($\Delta L/a$) of the junction. (Right panel) Repetition of the simulation presented in the left panel but now with $\Delta L = 0.85a = 0.0227\lambda_0$. **d**, (left panel) numerical results (black, green and blue) showing the magnitude of the transmission coefficient when the angle of between two stubs is 0°, 25° and 45°, respectively, considering stubs with $L_s = 0.5237\lambda_0$. The theoretical values (dashed red) for $L_s = 0.5\lambda_0$ (i.e. no added length as the theoretical transmission coefficient does not vary with angle between the stubs) are also plotted. **d**, (second panel) RMSE between the numerical and normalized ideal (linear V shaped) transmission coefficients for structures with angles between stubs ranging from −60° to 60°. **d**, (third panel) magnitude of the transmission coefficient for an angle between stubs of 25° as presented in the left panel when the length of the rotated stub is increased from 0.2 mm to 0.6 mm. The spectrum of the theoretical/ideal case (dashed red) is shown for completeness. **d**, (right panel) the calculated additional length of the stub required to minimize the RMSE between the numerical and ideal transmission coefficients for stub angles ranging from −60° to 60°.

**First order temporal differentiator: Transmission and reflection operation modes.**

To further study the performance of the proposed structures for temporal differentiation, full-wave numerical simulations were carried out using the time-domain solver of the commercial software CST Studio Suite®. A full description of the simulation setup is presented in the method section. A first order temporal differentiator was modeled using two identical closed-ended stubs with the same parameters as those used in Fig. 2d, with all the waveguides having a cross-section with dimensions $w = h = 1$ mm ($0.0267\lambda_0$, with again $\lambda_0$ as the wavelength in free space at 8 GHz) and being filled with vacuum ($\varepsilon_r = 1, \mu_r = 1$). In this section, the performance of the differentiator is evaluated working in both transmission and reflection configurations. From TL theory, it is expected that the transmission and reflection coefficients of the designed 4-waveguide structure will be complementary (provided that losses are negligible[44]). Based on this, while the V-shaped spectrum of the transmission/reflection coefficient will have their minimum at different frequencies, the width of the linear spectral region in the transmission and reflection coefficient will be the same (see insets of Fig. 3a,c, respectively). As in the right-most panel of Fig. 2c, the length of the two stubs were chosen to be $L_s = 19.637$ mm ($0.5237\lambda_0$), in order to produce a V-shaped dip of the transfer function at 8 GHz or 4 GHz when working in transmission or reflection mode, respectively (see Fig. 3a and Fig. 3c, respectively). From this design, it will be expected that when an incident temporal signal is applied from the input waveguide with a modulation frequency of 8 GHz, the differentiated signal in the time domain will be observed at the output waveguide while the temporal differentiated signal will appear at the incident waveguide (reflected signal) when the modulated frequency is 4 GHz (see the corresponding transfer function of each transmission/reflection configurations as insets in Fig. 3a,c).

To verify this, the numerical results of an incident temporal signal having a Gaussian envelope ($\sigma = 0.5$ ns with a maximum voltage of 1 V) modulated at 8 GHz and 4 GHz are shown in Fig. 3a,b and Fig. 3c,d (see incident signal on the left panel from Fig. 3b,d). To excite the structure, a waveguide port is used on the input waveguide (called port 1) and the results in transmission mode are recorded using a second port at the end of the output waveguide (port 2). With this configuration, the recorded time domain voltage at port 2 is shown in the middle panel of Fig. 3b (blue line) along with the theoretically calculated temporal derivative of the envelope of the incident signal (dashed-red line). Finally, the frequency spectra for both numerical and theoretical results are also shown in the right panel of the same

Fig. 3b. As observed an excellent agreement is obtained in both the time and frequency domain. For completeness, the numerical results of the space-time propagation of the incident signal is shown in Fig. 3a (calculated at $x = y = 0$ along the $z$-axis) corroborating how the transmitted signal corresponds to the temporal derivative of the incident signal. Similarly, the results of the structure working in reflection are shown in Fig. 3c-d where the incident and reflected signals in the time domain are shown on the left and middle panels of Fig. 3d along with the spectral content of the reflected signal in right panel of the same figure. By comparing the numerical and theoretical results in reflection mode, one can notice an excellent agreement between them. However, note how the spectrum of the numerical results (green line from the right panel from Fig. 3d) is not symmetric with higher frequencies having smaller amplitudes. As discussed in the previous section and shown in Fig. 2c, due to the non-zero value of the waveguide cross-sections, the length of the stubs should be tuned so that the perfect splitting at the junction happens at the required target frequency. Following this approach, the structure studied in Fig. 3 is tuned to operate at 8 GHz in transmission (with an added length of the stubs of $0.0237\lambda_0$). Hence, it is expected to obtain a slight deviation of frequency for the theoretical minimum of the V-shape of the reflection coefficient as it occurs at a different frequency (theoretically at 4 GHz). In the case shown in Fig. 3c,d, the central frequency of the reflection coefficient in the simulation is 3.904 GHz, which is slightly deviated from 4 GHz, as expected, producing an asymmetric reflection coefficient as shown in Fig. 3d. The space-time diagram is shown in Fig. 3c when working in reflection, demonstrating how the reflected signal still corresponds to the temporal derivative.

For completeness and to demonstrate that the designed structure can work with different incident signals, an arbitrary incident signal was also implemented. Here, the incident signal was defined by converting the profile of a landmark from Newcastle Upon Tyne, the Tyne Bridge, into a time domain signal as shown in Fig. 3e. The resulting signal after passing through the proposed structure is shown in Fig. 3f. By comparing the numerical output and the theoretical derivative found via the finite difference method, it can be seen how the proposed temporal differentiator can successfully identify the location of the edges in the structure of the bridge (denoted by the peaks in the derivative) as well as calculating the value of the slope along the arc of the bridge. These results demonstrate how the proposed structure can be used for *edge detection* of temporal signals.

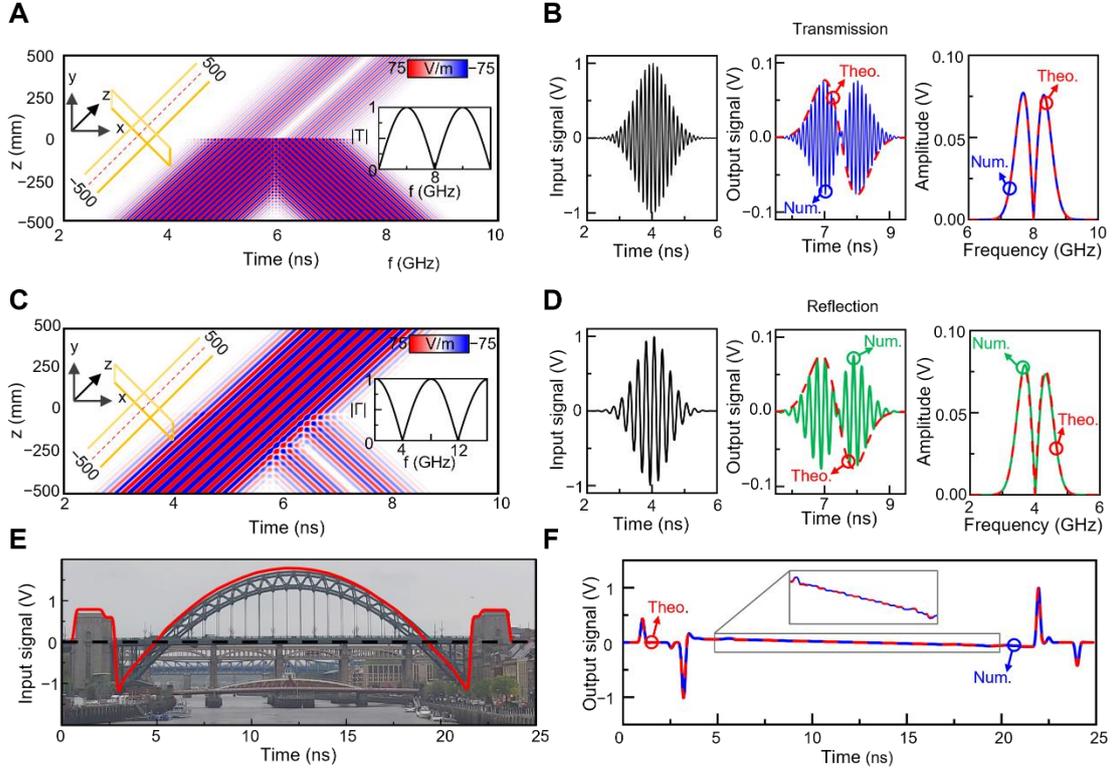

**Fig. 3| First order numerical results.** Numerical results of a first order differentiator with two metallic terminated stubs of length $0.5237\lambda_0$ made from waveguides with cross-section $w = h = 0.0267\lambda_0$ ($\lambda_0 = 37.5$ mm, $f_0 = 8$ GHz). **a**, Numerical results of the electric field distribution in space and time for our proposed first order differentiator considering an incident 8 GHz modulated Gaussian (standard deviation $\sigma = 0.5$ ns). These results are calculated along the propagation axis of the whole structure at $x = y = 0$. **b**, Time domain simulation results of the scenario presented in **a**, calculated at the ends of the input and output waveguides ($z = \pm 500$ mm $= 13.3\lambda_0$). The input signal in the time domain is shown on the left panel as a black line, along with the numerical results of the recorded voltage (middle panel) calculated at the end of the output waveguide ($z = 500$ mm $= 13.3\lambda_0$) and the theoretical derivative of the envelope in the time domain (blue and dashed red line respectively). The frequency content of the incident and output signals is shown on the right panel for completeness. **c, d**, same as panels **a, b**, considering the same structure but working in reflection configuration. Here, we use a 4 GHz modulated incident Gaussian signal (same standard deviation as **b**). The numerical results of the reflected signals both in the time and frequency domain are plotted as green lines in **d**. **e**, Unmodulated arbitrary input signal representing the Tyne Bridge (red line), a local landmark from Newcastle Upon Tyne in the United Kingdom. **f**, Numerical (blue line) and theoretical (dashed red line) results of the output voltage as a function of time for the scenario from **e**.

**Temporal differentiation of $m^{th}$ order.**

As demonstrated in the previous sections, first order differentiation can be performed by an individual block emulating this operation using, in our case, interconnected TLs (see Fig. 1a,b, Fig. 2,3). Higher order differentiation can be achieved by, for instance, cascading multiple first order differentiators together such that the first order operation is performed onto the incident wavefront multiple times in series. In other words, the ideal transfer function of an $m^{th}$ order differentiator can be found by multiplying the first order transfer function $m$ times, i.e., it can be mathematically represented by defining a transfer function, as follows:

$$T_m = [2\pi i (f - f_0)]^m \qquad (5)$$

Interestingly, this transfer function also holds for fractional derivatives given that the order $m$ of the derivative can be a non-integer value[50]. In the time domain, these fractional derivatives can be found by using, for instance, the Riemann-Liouville equation[51].

$$\frac{\partial^m f(t)}{\partial t^m}\Big|_{t>b} = \frac{1}{\Gamma(\lceil m \rceil - m)} \frac{d}{dt^{\lceil m \rceil}} \int_b^t (t-x)^{\lceil m \rceil - m - 1} f(x) dx \qquad (6)$$

where $\Gamma$ is the gamma function[52] (a function commonly used to extend factorials into complex numbers[53]), $\lceil m \rceil$ denotes rounding $m$ upwards to the next integer, $t$ is the variable which the function $f(t)$ is being differentiated with respect to, $x$ is a substitute variable used to calculate the integral and $b$ is the basepoint of the system which describes the non-locality of non-integer derivatives[54].

To carry out this operation using the proposed waveguide junctions, here we present a general structure for $m^{th}$ order differentiators as schematically shown in Fig. 4a. It consists of multiple "layers" of cascaded first order differentiators being connected via parallel plate waveguides as TLs. The number, length and open/closed nature of the stubs can all be individually defined for each differentiator block, meaning that each differentiator does not necessarily need to be the same as its adjacent blocks. The length between each differentiator can also be defined, allowing for a higher degree of control over the spectrum of the transfer function of the full structure. Note that such control is particularly important when considering interconnected blocks of TLs as multiple reflections between blocks need to be considered and can be indeed tuned at will by exploiting all the different parameters within the full structure. In this context, as it is known in filter design, connecting layers together by a TL of length $\lambda_0/4$ will increase the "order" of the differentiator, increasing the bandwidth of the filter (an increased bandwidth of the minimum in the transmission coefficient in our case[44,55]). This can be understood as the multiplication of the transmission coefficients of the individual layers in the region around $f_0$. As differentiation is performed around a minimum in the transmission coefficient, when considering lossless TLs, the majority of an incident signal will be reflected by the differentiator. Based on this, when cascading multiple differentiators, the high reflection coefficient of each individual differentiator will produce a large standing wave between the layers. A distance of $\lambda_0/4$ is then chosen to connect the different layers (differential operators) to ensure that the reflections between layers will destructively interfere with one another thus not impacting the output of the subsequent differentiator. Hence, by using TL theory, choosing the length of the waveguide connecting differentiators as an odd integer multiple of $\lambda_0/4$ will preserve the symmetry of the transmission coefficient around the modulation frequency $f_0$ (a symmetry requirement due to the nature of Eq. 5 around $f_0$). An in-depth study of the response of the $m^{th}$ order differentiator when the length of the waveguides connecting the differentiators is included in the supplementary materials section S3 for completeness[56].

As discussed above, reflections between cascaded layers are expected to be large, hence approximations such as the theory of small reflections[44] cannot be used. Instead, we utilize the Redheffer star product[56,57] method to calculate the transfer functions of the cascaded structures. This method is an alternative to the commonly used transfer matrix method (TMM)[58], which enables the symmetry of the scattering matrix to be exploited for greater computational efficiency. This method can be briefly explained as follows:

Consider a pair of scattering matrices $S^1$ and $S^2$ connected together such that the output of one matrix feeds into the input of the other, and vice versa (see schematic representation in the bottom panel of Fig. 4a). Mathematically, this can be expressed as follows:

$$\begin{pmatrix} y_{1L} \\ y_{1R} \end{pmatrix} = \begin{pmatrix} S^1_{11} & S^1_{12} \\ S^1_{21} & S^1_{22} \end{pmatrix} \begin{pmatrix} x_{1L} \\ x_{1R} \end{pmatrix}, \quad \begin{pmatrix} y_{2L} \\ y_{2R} \end{pmatrix} = \begin{pmatrix} S^2_{11} & S^2_{12} \\ S^2_{21} & S^2_{22} \end{pmatrix} \begin{pmatrix} x_{2L} \\ x_{2R} \end{pmatrix} \quad (7a)$$

$$x_{1R} = y_{2L}, \ x_{2L} = y_{1R} \quad (7b)$$

where the $S^1_{oi}$ and $S^2_{oi}$ terms are the scattering coefficients of the first and second scatter (as labeled by the numbered superscript), respectively, $o$ and $i$ represent the output and input waveguides which the scattering coefficient relates to, respectively. The $y$ and $x$ terms are the output and inputs of each scatterer, respectively. The numbered subscript denotes which scatterer the input corresponds to (1 meaning the first and 2 meaning the second scatterer, respectively) while the subscripts $L$ and $R$ represent where the output/input is taken (left and right of the waveguide junction, respectively). Based on this, a combined scattering matrix $S^3$ can be written

$$\begin{pmatrix} y_{1L} \\ y_{2R} \end{pmatrix} = \begin{pmatrix} S_{11}^3 & S_{12}^3 \\ S_{21}^3 & S_{22}^3 \end{pmatrix} \begin{pmatrix} x_{1L} \\ x_{2R} \end{pmatrix} \qquad (8)$$

where the $S_{oi}^3$ terms are the scattering coefficients of the overall structure, defined as:

$$S_{12}^3 = S_{12}^1(1 - S_{11}^2 S_{22}^1)^{-1} S_{12}^2 \qquad (9a)$$

$$S_{11}^3 = S_{12}^1(1 - S_{11}^2 S_{22}^1)^{-1} S_{11}^2 S_{21}^1 + S_{11}^1 \qquad (9b)$$

$$S_{21}^3 = S_{21}^2(1 - S_{22}^1 S_{11}^2)^{-1} S_{21}^1 \qquad (9c)$$

$$S_{22}^3 = S_{22}^2 + S_{21}^2(1 - S_{22}^1 S_{11}^2)^{-1} S_{22}^1 S_{12}^2 \qquad (9d)$$

In general, the $S_{oi}^3$ terms in Eqs. 7-9 can be written as matrices representing the scattering between multiple ports in a network, however this is not necessary for our implementation as we consider scatterers with only two ports (one input, one output). As such, the calculated $S_{oi}^3$ terms represent the transmission and reflection coefficients for a signal from input $i$ towards output $o$ of the structure (i.e. $S_{21}^3$ and $S_{11}^3$ is the transmission and reflection coefficient of the full structure which is the result of the combination of two scattering matrices together, respectively, when applying the incident signal from the left). Note that in this configuration, the effect of the connecting length is absorbed into one of the scatterers (e.g. $\boldsymbol{S^1}$) by adding a phase change to the transmission and reflection coefficients from the connecting waveguide (i.e. $S_{21}^1 \rightarrow S_{21}^1 e^{-i\varphi}, S_{22}^1 \rightarrow S_{22}^1 e^{-2i\varphi}$, where $\varphi = \omega L_c \sqrt{\varepsilon_r \mu_r}/c$ is the electrical length of connection), $\omega$ is the angular frequency of the signal and $L_c$ is the length of waveguide connecting the two layers. Moreover, due to the reciprocal nature of our proposed differentiators (individual layer and overall $m^{th}$ order structure), it is expected that $S_{11}^3 = S_{22}^3$ and $S_{12}^3 = S_{21}^3$, therefore only two calculations are necessary to combine adjacent layers. Finally, the Redheffer star product is the operation which combines the matrices in Eq. 7a into the matrix in Eq. 8 using the relations in Eq. 9. This can be written as[59]

$$\boldsymbol{S^3} = \boldsymbol{S^1} \star \boldsymbol{S^2} \qquad (10)$$

with "$\star$" representing the star product. From this, the scattering matrix of the cascaded system is found by repeatedly combining the scattering matrices of adjacent junctions until all junctions have been encapsulated into a combined scattering matrix. The transmission and reflection coefficients are then found by taking the $S_{21}^3$ and $S_{11}^3$ terms from the combined scattering matrix, respectively.

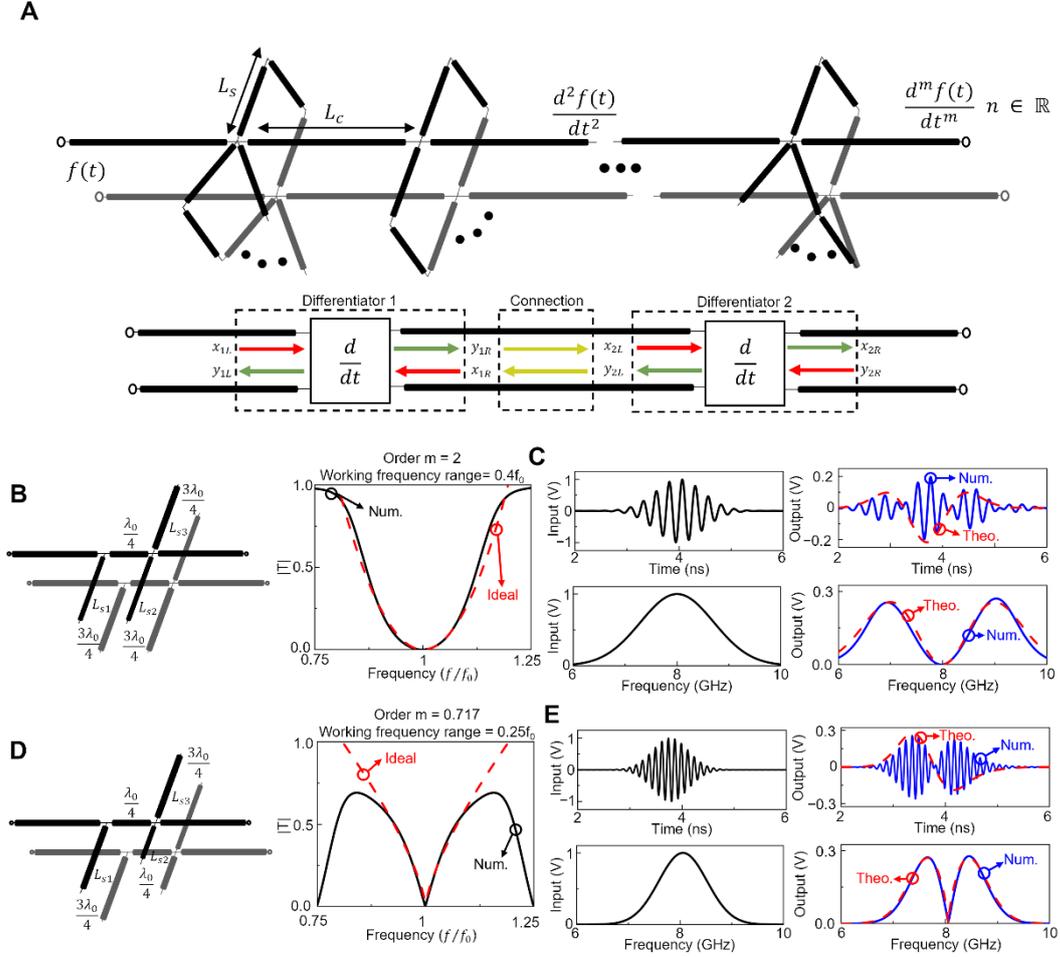

**Fig. 4| Arbitrary $m^{th}$ order derivatives. a**, (Top panel) Schematic of the proposed approach showing cascaded differentiators. The number of stubs, its length ($L_s$), open/closed nature, and length of the waveguides used to connect two junctions of waveguides ($L_c$) may be controlled to tailor the desired transfer function. (Bottom panel) Block diagram of two interconnected differentiators. Inputs (red arrows) and outputs (green arrows) of the individual differentiators are presented along with the reflections produced between the two differentiation blocks (yellow arrows). **b**, **c**, Example of a second order differentiator made from two interconnected junctions. **b**, TL representation (left) along with the ideal/theoretical (red) and numerically calculated (black) magnitude of the transmission coefficient (right panel). The waveguides have a cross-section of $w = h = 0.0267\lambda_0$ and dimensions $L_{s1} = L_{s2} = L_{s3} = 0.7596\lambda_0$ **c**, Time domain simulation of an 8 GHz central frequency Gaussian (standard deviation $\sigma = 0.46$ ns) incident signal in the time domain (top-left) and its corresponding spectrum (bottom-left) along with the numerical (blue) and theoretical values (dashed-red) in the time domain (top-right) and frequency domain (bottom-right). **d**, **e**, The same as panels **b**, **c**, but for a design to perform fractional differentiator of order $m = 0.717$. Here, the incident signal has a different time duration (standard deviation $\sigma = 0.3536$ ns) to fit its spectrum within the working frequency range of $0.25f_0$. The stub waveguides have dimensions $L_{s1} = L_{s3} = 0.758\lambda_0, L_{s2} = 0.505\lambda_0$

With this method we cycle through various possible designs, at each stage evaluating the quality of the differentiator by calculating the RMSE between the calculated and ideal transfer functions. The design which best matched the desired transfer function was then modeled and simulated in CST Studio Suite® to evaluate its performance in a full-wave simulation software. To test the flexibility of this method we designed and evaluated two further devices. The first, shown in the left panel of Fig. 4b, was designed to perform second order differentiation. This requires a quadratic transfer function resembling $[2\pi i(f - f_0)]^2$ as $m = 2$ in Eq. 5. The ideal and numerical transfer functions (transmission coefficient) can be found in the right panel of Fig. 4b. The working frequency range (here defined as the spectral range around $f_0$ before the numerical transfer function deviates from the ideal spectra by 10%) was found to be $0.4f_0$. To evaluate the designed structure, a Gaussian signal in the time domain modulated at 8 GHz was used as the excitation signal (see Fig. 4c). The standard deviation of this signal is $\sigma = 0.3536$ ns, chosen such that the spectrum of the incident signal would fall within the working frequency range of the differentiator. The numerical results of the time domain signal calculated at the output of the structure is shown in Fig. 4c (blue line) along with the theoretical derivative of the envelope (red dashed line) corroborating how it is possible to calculate the second order derivative with the designed structure. As can be observed in Fig. 4c there are small variations between the calculated and ideal derivative. This is explained by the small mismatch between the ideal and numerical transfer functions at frequencies farther from $f_0$ as seen in Fig. 3c. This is an expected result as the optimization weighed differences between the ideal and numerical transmission coefficient higher in the region around $f_0$ when calculating RMSE.

For completeness, and to demonstrate that the order of the temporal derivative does not necessarily need to be an integer, a structure with the ability to perform the fractional derivative of order $m = 0.717$ (randomly chosen) was also designed. The design and transmission coefficient of this structure are shown in Fig. 4d. The working frequency range around $f_0$ in which the transmission coefficient resembled the ideal curve for the corresponding order $m = 0.717$ was found to be approximately $0.25f_0$ (calculated as described above). This can be seen in the right panel from Fig. 4d where the numerically calculated transmission coefficient (black plot) agrees with the ideal transfer function (red line) within a certain frequency region around $f_0$ but it diverges at larger and smaller frequencies. As before, a time domain simulation with an incident modulated (8 GHz) Gaussian ($\sigma = 0.4632$ ns, so that its spectrum will fall within the working frequency range, $f_0 \pm 0.25f_0$) was carried out to evaluate the performance of the fractional temporal differentiator. The numerical results of the calculated output voltage (blue plot) in both time and frequency domains are presented in Fig. 4e where it is clear how, by comparison to the theoretical value (red plot), the output signal represents the fractional temporal derivative of the incident temporal signal. As these results show, the envelope of the output signal in the time domain has two "lobes" which are asymmetrical around the central dip (i.e., the lobes have different amplitudes and temporal duration), compared to the first order differentiation case presented in Fig. 3. This asymmetric temporal signal is an expected feature of fractional derivatives with $0 < m < 1$ of Gaussian signals[60,61]. These results demonstrate that the proposed structure indeed has the ability to perform fractional differentiation onto the incident temporal signal.

**Discussion or Conclusions**

In summary, a method for performing analogue differentiation to the envelope of incident temporal signals has been proposed by exploiting the splitting and superposition of TEM waves within parallel plate waveguide junctions. To do this, close-ended stubs connected at such junctions were used to tailor the transmission coefficient of the proposed structure to resemble the $m^{th}$ order differentiation operator in the frequency domain. A full mathematical description of the splitting and superposition of signals within these structures has been presented in terms of the scattering matrix approach. Different designs have been demonstrated numerically such as the calculation of first ($m = 1$) and fractional order ($m = 0.717$) temporal differentiation of a temporal Gaussian envelope sinusoidally modulated (modulation frequency 8 GHz). Additional examples included envelopes of arbitrary shapes and the use of the technique in reflection and transmission mode, among other studies. A good agreement was found between all the presented results and their corresponding analytic calculations. We envision that this work may enable the development of further time domain wave-based analogue computing devices opening new directions for high-speed computing.

**Methods**

The numerical simulations shown in Fig. 1c, Fig. 2, Fig. 4b and Fig. 4d were performed using the frequency domain solver of the commercial software CST Studio Suite® while the results in Fig. 1e, Fig. 1f, Fig. 2, Fig. 4c and Fig. 4e with the time domain solver. Parallel plate waveguides (top and bottom PEC boundary conditions) with a cross section ($w = h = 1$ mm $= 0.0267\lambda_0$, where $\lambda_0 = 37.5$ mm) where implemented, unless stated otherwise in the main text. Vacuum ($\varepsilon_r = 1, \mu_r = 1$) was used as both the waveguide filling material and the background medium of the simulation. Waveguide ports were used to excite/extract the input/output signals. These ports were placed at the ends of the input/output waveguides with the latter having a length of 25 mm ($0.667\lambda_0$) from the ports to the position of the junction. For the results shown in Fig. 3a,c, this separation was instead 500 mm = $13.3\lambda_0$ to better observe the waves propagating in the spacetime diagrams. Boundary conditions were set to *open (add space)* in the $x$ and $y$ axis and to *open* in the $z$ dimension to add background space after the structure and to avoid undesirable reflections, respectively. Gaussian signals in the time domain simulations were defined following the equations $G(t) = e^{-(t-4)^2/2\sigma^2} \sin(2\pi f_0 t)$, where $f_0$ is the modulation frequency, $\sigma$ is the time domain standard deviation and $t$ is time.

## Acknowledgements

V.P-P. and A. Y. would like to thank the support of the Leverhulme Trust under the Leverhulme Trust Research Project Grant scheme (RPG-2020-316). V.P.-P. also acknowledges support from Newcastle University (Newcastle University Research Fellowship). V.P-P. and R.G.M would like to thank the support from the Engineering and Physical Sciences Research Council (EPSRC) under the scheme EPSRC DTP PhD scheme (EP/T517914/1).

## Conflicts of interests

The authors declare no conflicts of interests.